# Continuous Flow Analysis to Detect Security Problems

Steven P. Reiss
Department of Computer Science
Brown University
Providence, RI
spr@cs.brown.edu

*Abstract*—We introduce a tool that supports continuous flow analysis in order to detect security problems as the user edits. The tool uses abstract interpretation over both byte codes and abstract syntax trees to trace the flow of both type annotations and system states from their sources to security problems. The flow analysis achieves a balance between performance and accuracy in order to detect security vulnerabilities within seconds, and uses incremental update to provide immediate feedback to the programmer. Resource files are used to specify the specific security constraints of an application and to tune the analysis. The system can also provide detailed information to the programmer as to why it flagged a particular problem. The tool is integrated into the Code Bubbles development environment.

## I. INTRODUCTION

Insecure applications at all levels have become a serious problem for individuals, businesses, and our national security. The best way to have secure applications is to write them securely in the first place. The best way to maintain secure applications is to ensure that all changes maintain the original level of security and new security vulnerabilities are fixed immediately. Our aim is to provide programmers with the tools they need to detect and fix security problems as they write and maintain code. These tools provide immediate feedback to the programmer on security problems while they are editing, much as current development environments provide immediate feedback on syntactic problems.

The CHEETAH system provided an initial framework for offering feedback on flow problems when the user rebuilds on flow-related problems [11]. It offers a just-in-time strategy that gives immediate feedback at a course level within seconds, and then provides more detailed feedback over time. While it defines a general strategy for converting a batch analysis into an incremental analysis, it was only used with taint analysis. However, experiments with CHEETAH showed that it could be beneficial and that programmers liked the feedback.

In this paper we show that it is possible to do a more comprehensive flow analysis for a variety of security problems that can provide complete feedback within seconds without having to resort to just-in-time methods. Our analysis is incremental, multi-threaded, and can be specialized to the application at hand, a necessity for security checking. It is capable of handling relatively large (100's of KLOC including libraries) Java systems, and is suitable for integrating into a development environment. Although our initial analysis may take up to a minute, we can provide incremental feedback within seconds on moderate sized systems, and in under a second on smaller systems. This makes the analysis fast enough so that it can be run as the user types rather than just when they rebuild, providing true immediate feedback. Finally, the system can be specialized to the particular security concerns of a system.

Most security checking tools today, for example Microsoft PREfix, HP Fortify, or Coverity, run off-line. They use a variety of different analyses to detect potential problems over a complete system, either source, binary, or a combination of the two. While these tools are useful, they can be overwhelming to the programmer. FindBugs [15], for example, running on a large system, can return thousands of potential problems. Programmers can not deal easily with such output and tend to either ignore it, restrict the tool to only find "known" issues, not all potential issues, thereby missing some problems, or run the tool infrequently [19]. Other tools, while providing more detailed or accurate analysis, take significant time to process a complex system. Programmers are not willing to spend this effort after every minor change and hence the tools do not always get run appropriately. This is the motivation both for CHEETAH and for our approach.

Our analysis tracks the sources and types of values that can reach each point in the program. The type information, based on Java 8 type annotations, includes annotation-based subtypes such as *HtmlTainted* or *SqlUntainted*. The analysis also maintains global type-state information such as *User_Authenticated* or *Role_Administrator* for each program point. Customization of the analysis permits the rapid addition of specialized security checks to address new or uncovered vulnerabilities, the ability to describe the specific security constraints of the current application, and the ability to adapt the analysis to different frameworks and libraries.

The analysis framework is integrated into the Code Bubbles development environment [3] demonstrating its feasibility. The analysis, once started by the environment, monitors the programmer's current working set and edits, updates the analysis incrementally after edits, and provides asynchronous feedback with the results. It also supports programmer queries to provide more information about a potential error.

The contributions of this paper include:
- An architecture for a complete flow analysis, FAIT, that is fast enough to run as the user edits with enough accuracy to detect potential security problems;
- Resource files for customizing the analysis and making it fast and accurate;
- Techniques for mining the flow analysis to provide the user with a detailed description of a potential problem; and
- An initial integration of this tool into an integrated development environment.

The remainder of the paper starts with an overview of related work in Section II. The way we view security problems is specified in Section III. A description of the flow analysis framework is given in Section IV. Specialization of the analysis is discussed in Section V. Incremental update is described in Section VI. Mining the analysis to provide feedback is detailed in Section VII. The integration into Code Bubbles is described in Section VIII. Finally, an initial evaluation of the approach is given in Section IX.

## II. RELATED WORK

Type annotations were proposed as a means for doing better static analysis of Java programs [10]. They are now part of the Java language, although tools that make use of them are not part of the distribution. The most common tool is the Checker Framework [6], which runs as a separate process or as an Eclipse plug-in, on demand by the user.

Type annotations can be difficult to use [30]. For most type annotations to be effective, a potentially large number of annotations might need to be added to the program. For example, the recommended way of using @*Tainted* annotations [12] is to annotate *all* the locations in the program where there could be tainted data [6]. This can be a lot of work, and work that would have to be updated continually as the program evolves. Flow analysis does an implicit type analysis in that it determines the types of data that can flow to each point in the program and does not require all the intermediate type annotations. Using flow analysis rather than type analysis can simplify what the programmer needs to do while still providing a reasonable level of security.

Our flow analysis builds on large-scale analysis libraries such as Soot [45] and Wala [17]. Our work on making FAIT incremental is based on work in the areas of incremental execution [24,32] and continuous and incremental program analysis [1,29,37,46]. Converting FAIT to multi-threading built on past work in this area [21]. Recent work has demonstrated that some program analysis, for example pointer analysis, can be done quite efficiently incrementally [47]. Cobra provides an example of a fast query interface for a flow analysis similar to what we to provide [14].

Flow analyzers that attempt to handle larger systems often have special means for handling libraries and native methods including summarization of a library or a component [36], on modeling of a library method through either procedural code [23] or a specialized model based on the type of analysis [38]; on techniques for handling dynamic class loading [41]; and on techniques for handling callbacks [36]. The notion of performance analysis as applied to program synthesizers [2] can also be applied to flow analyzers. The resource files used by FAIT allow modeling library methods, handling dynamic loading and other instance of Java reflection, and allow performance tuning of the analysis.

The idea of using flow information while editing comes from early work by Zadeck [46] which only addressed use-def linkages that are handled by today's compilers. Cascade infers type annotations [44] based in part on the principle of speculative analysis [5], i.e. showing programmers the consequences of their actions ahead of time, which is one of our goals as well. This work led to an on-line form of some of these analysis which demonstrated its benefits [28]. Just-in-Time analysis as in CHEETAH provides a compromise, with relatively fast but specialized analysis that gets more detailed as the programmer waits [11]. Our efforts aim to go beyond this by doing a complete analysis within seconds.

Safety conditions related to control flow rather than data flow require a specification beyond type annotations. Such conditions are more naturally represented as automata as in Flavers [8] or CHET [33]. An alternative is to use temporal logics such as LTL or CTL [7]. Within a data flow analysis, these can be viewed as roughly equivalent, assigning a set of possible states to each location in the program and updating the sets based on the flow analysis which is what FAIT does.

Many tools have been proposed for off-line security checking via static or dynamic analysis. While we have mentioned the more general tools like PREfix, Fortify, Coverity, FindBugs and Checker, there have also been numerous specialized efforts. Shahriar and Zulkernine [39] and Siponen and Oinas-Kukkonen [40] provide overviews of these. Lam, et al. used queries over a deductive database as a lightweight means to identify security problems [20,26,27]. Recent efforts along these lines include JSPChecker [42], Andromeda [43], SFlow [16], SymJS [22], Pidgen [18], and AndroidLeaks [13]. Our efforts show that these off-line analyses can be done on-line and the various specialized techniques for providing early feedback are not necessary.

## III. SPECIFYING SECURITY PROBLEMS

In order for security checking to be practical within a programming environment it must be easy to define and specify the security properties of each application. This means making it simple and practical to define type-based security constraints as well as state or automata based constraints. For each, the developer needs to define the underlying properties and annotate the program.

Defining security constraints independent of an application will not provide a complete or correct analysis. Security issues depend on the application, its libraries, and the actual code involved. For example, in checking for cross-site scripting issues, the incoming HTML can be sanitized before it is put into a database or when it is taken from the database. These two implementations require different sets of annotations. Moreover, different applications might require different types of HTML sanitization, allowing different sets of tags to persist. The set of authentication roles is also application dependent.

For type-based constraints, one needs to define the set of type states the constraint can use, how these are defined by type annotations, and how different operators can affect them. Checker [6] does this by having the user write code. Our approach is to define all the information in the resource file, with each annotation requiring 50 or fewer lines of XML. This involves defining the rules for merging two values, for restricting a value to a particular subtype, and for handling operators. It also specifies when errors should be output. We = provide a user interface to make this definition more convenient. One difficulty is that the associated annotations need to be compiled and available when compiling the source. We provide a generic *@Secure(...)* annotation that can be used, but it is not as convenient as direct annotations.

Using annotations with flow analysis is generally simpler than using them with type analysis. One only needs to annotate the sources and possible bad sinks for the flow, but does not have to annotate all the intermediate locations. As an example, we have taken an insecure web server developed for one of our courses as a security exercise (based on a version of the OWASP NodeGoat project [31] modified to use SQL). and converted it into a Java-based server using *nanohttpd*. To detect SQL injection attacks, we marked the input from the user as *SqlTainted* using type annotations. We needed to do this for the three input sources (request data, URL data, and cookies). Then we annotated the query parameter on the SQL call as *SqlUntainted*. A hidden annotation type, *MaybeSqlTainted*, was used internally to track other data. A fourth annotation type, *SqlFullyTainted*, is used to denote data that cannot be untainted safely (e.g. by an *sanitize* routine). The type propagation rules indicated that any operation with tainted data yielded tainted data, that concatenating a tainted string yields a fully tainted one, and operations with only untainted data (which includes all constants and initially all allocated items) yielded an untainted result. With this model, adding these four annotations was sufficient to detect the various ways that the server was vulnerable to SQL injection attacks.

To handle objects that have contents, for example arrays or lists or maps, there are two choices. One can introduce additional annotations to indicate the content of an object is or is not tainted, or one can actually apply any annotations on the object to its contents. While the Checker framework does the former, we decided to do the latter. Our experience showed that this was easier for the programmer, required fewer annotations, and did not affect what the programmer wanted to specify since it would be rare to distinguish between the two.

Automata-based checks are useful for control-flow based security problems such as those involving authorization or access control. Our approach is to define events in the source code that can be used to trigger state transitions in the automaton. Automata are defined in the resource files as a set of states and transitions based on named events, with some transitions flagged as generating an error. A user interface is provided to aid the user in this definition. Events currently are defined in the program by calls to an empty routine *KarmaEvent( "event_name")* from the Karma library. This call can be used in an assert statement for clarity.

As an example, we used an automaton to check the authentication logic in our insecure web server. To do this, we created a simple automaton with states for NONE, USER, ADMIN, and ERROR indicating the known and verified user role at each point in the code. This was defined from about 30 lines of XML. We added event indications at the start of URL processing, where the user was authenticated, and where administrative access was checked. We then added event indications at each entry point in the code where we assumed that the current role should be user or administrative. This yielded a total of seven event indications. This was then sufficient to find the problematic path whereby authentication could be avoided and to validate that other paths were correct.

## IV. A Fast and Accurate Flow Analysis

Checking type annotations and automata-based safety conditions on-line requires a flow analysis that is fast, relatively complete, and accurate enough to detect potential problems. Our system meets this requirement and, with incremental updates, is capable of providing feedback to programmers as they type. The system offers a practical balance between accuracy and performance, includes means for handling complex programs and libraries, and ensures that the various security properties can be checked.

When the system is used with the insecure web server described above, it computes the initial set of potential security problems in about 10 seconds, highlighting the error locations in any editor open and in a separate window describing security issues. After that, as the user edits, the set of problems (and their corresponding displays) are updated almost immediately (under 1 second). If the user attempts to correct one of these problems, they will know right away if their correction is effective.

**Abstract Interpretation.** FAIT does flow analysis using abstract interpretation [9]. The system interprets the program while maintaining an abstract state for each program point. Where execution points merge, the program states are merged. The analysis is thus flow-sensitive, but not path sensitive.

The system actually contains two abstract interpreters. A byte-code interpreter is used to process library files and those parts of the user's program that are not currently being edited. An abstract syntax tree interpreter is used to process files the user is currently viewing and might be editing. The decision of which interpreter to use is made at the start of each method.

The abstract-syntax tree interpreter uses the Eclipse parser to create abstract syntax trees and the $S^6$ semantic analyzer to do semantic binding [34]. It modifies the abstract syntax trees

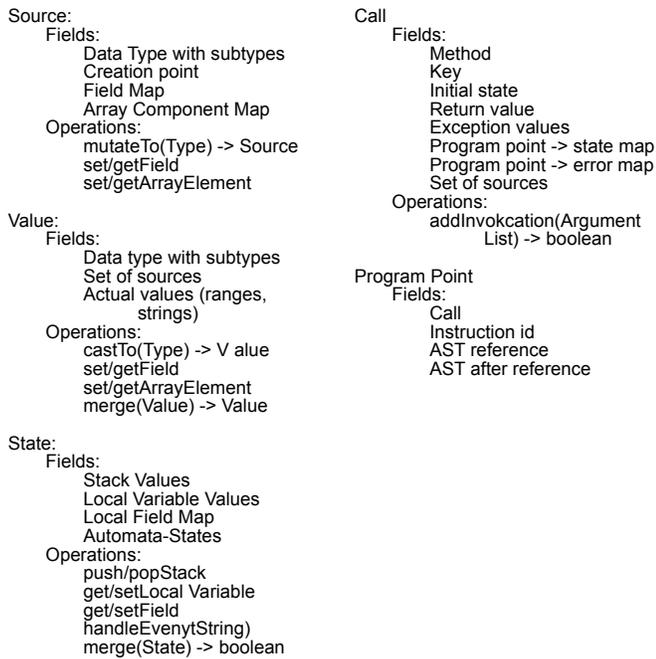

```
Source:
    Fields:
        Data Type with subtypes
        Creation point
        Field Map
        Array Component Map
    Operations:
        mutateTo(Type) -> Source
        set/getField
        set/getArrayElement
Value:
    Fields:
        Data type with subtypes
        Set of sources
        Actual values (ranges,
            strings)
    Operations:
        castTo(Type) -> Value
        set/getField
        set/getArrayElement
        merge(Value) -> Value
State:
    Fields:
        Stack Values
        Local Variable Values
        Local Field Map
        Automata-States
    Operations:
        push/popStack
        get/setLocal Variable
        get/setField
        handleEvenytString)
        merge(State) -> boolean

Call
    Fields:
        Method
        Key
        Initial state
        Return value
        Exception values
        Program point -> state map
        Program point -> error map
        Set of sources
    Operations:
        addInvokcation(Argument
            List) -> boolean
Program Point
    Fields:
        Call
        Instruction id
        AST reference
        AST after reference
```

Fig. 1. Basic flow analysis data structures

to move field initializations inside either a static initializer or constructors, to add default constructors, to add implicit calls to the super constructor, to define enumeration constants and implied enumeration methods, and to merge static initialization blocks into a single method. The interpreter handles implicit calls derived from boxing and unboxing, enhanced for statements, lambdas, catch and finally statements, and varargs.

**Basic Data Types.** The basic data types used by the interpreters are shown in Figure 1. These include sources, values, and states. A *Source* is a representation of a specific value that has a specific creation point in the application. The flow analysis computes, for each source, all the points in the program to which the source can flow. Sources that can represent objects contain information about the values of the fields of the corresponding objects. Sources that can represent arrays contain information about the values of array elements, both for specific indexes and in general. This yields an effective shape analysis for each object. Field values are also kept globally for each distinct field for those cases where the source might represent an arbitrary object and for static fields.

Special sources are created to represent arbitrary values that might be returned by system routines or that are needed otherwise. These can be fixed or mutable. Mutable sources are automatically transformed into a new source with a valid subtype of the original type where appropriate, for example at a cast or a method invocation, provided that the subtype is known to the system either a priori or through a class initialization. This is necessary, for example, to handle system methods that return interface types or arbitrary objects.

The basic operations on sources include getting and setting both fields and array elements as well as a mutation operator which either returns the current source if it is of the right type, returns a new source of the designated type if the current source is mutable and the types are compatible, or returns null.

Sources are grouped together and generalized by typed *Values*. Each value consists of the extended data type that includes the basic Java type along with the associated subtypes, a description of the set of valid actual values, and a set of sources. The data type represents the information that would be generated by a type analysis.

The set of valid actual values can indicate any value of the given type or can be more restrictive. Integer values maintain separate upper and lower bounds, where either can be a number or represent an unbounded value. Operations are defined so as to maintain these bounds. Floating point numbers are kept without bounds since flow analysis for security rarely depends on the value of such variables. String constants retain their value.

Values are considered immutable and unique, that is, two values with the same data type, same source set, and same actual value constraints are the same object.

The basic operations on values include setting and getting fields and array elements, casting the value to a particular type, and merging two values. Setting and getting field and array elements call the corresponding operator on each source; the cast operator computes a new set of sources by restricting the current set, either keeping, discarding, or mutating each source, and then builds a new value accordingly. The merge operation combines two values, taking the union of their source sets and their actual value constraints, and computing a new data type.

The abstract interpretation maintains the values associated with the stack, each local variable, field, and array contents of the application. Local variables and the stack are kept as part of a *State* structure, which is computed for each reachable program point. It also contains the values of any non-volatile fields of the *this*-object that were explicitly set as well as the set of automata states relevant to each defined automata-based security property.

The basic operations on a state include stack manipulations such as popping and pushing values; getting and setting local variables; and accessing fields. The latter operation checks if the base value for the field access is 'this', and if so, either includes the value in the state or returns the value associated with the state if it exists. There is an operation to handle automata transitions that takes an event name and updates the set of automata states associated with the state. The final operation is a merge which combines two states. This does a merge of the associated stack, local and field values and a union of the automata states of the two states. It returns true if anything changed in the merge.

The analysis decides, at each point where a method is called, whether to create a new instance of the method for evaluation or to reuse an existing instance. Such instances are kept in a *Call* structure. The call structure indicates the method, the initial program state which includes the argument values being passed in, the return value if it is known, the set of exception values that might be thrown by the call, and the mapping of locations to program states. Call structures also keep track of all the sources explicitly created by that call and the set of errors generated by type or automata conditions within the call. Keeping this information in the call allows for simpler incremental update.

At each method invocation, the system finds a set of call objects, one for each method that could be called at that point. The determination of a call object for a particular method is based on the method, its source location, and the set of values being passed as arguments. This process makes the analysis partially context-sensitive. By default, all non-static methods within the application proper have separate instances created for each distinct value of the *this* parameter, while all system and library methods have only a single instance. This is a compromise that provides a more accurate analysis for code that is more critical to analyze, while still being efficient. The programmer can change these defaults for particular methods or libraries through a resource file, either making the call non-sensitive, choosing to only look at the this parameter, or choosing to look at all the parameters. This information is represented in the key field of the call object.

Call objects support an operation that initiates an invocation with a new set of arguments. This merges the argument values with any prior values in the initial state and returns true if this changes the initial state and thus the call should be reevaluated.

*Program point*s indicate locations inside a call. For analyzing byte code, they simply refer to the particular instruction of the method. For source code, they contain two references to abstract syntax tree nodes. The first is the node being interpreted. The second is the child of that node that just finished being interpreted, which is null when the node is first evaluated. This provides a level of detail for source interpretation that is roughly equivalent to byte code instructions.

**Analysis Algorithms.** The system maintains a work queue of calls where the flow needs to be analyzed. For each call in the work queue, it maintains a set of program points where that analysis should start. The analysis is done one call at a time, taking calls off this work queue. This approach allows the relatively easy use of multiple threads, with a pool of worker threads and a queue of methods to evaluate. The system uses a priority queue, with static initializers being given priority over constructors which have higher priority than normal methods. Methods called from initializers or constructors inherit their caller's priority.

The analysis starts with each main method in the application and its entry point. The analysis also takes into account JUnit test cases. Our experience showed that this provided better coverage, especially for libraries or packages that are used externally, When loading an application, it scans all project classes for code annotated with the *@Test* annotation and creates a dummy file with a main program that executes

```
AnalyzeCall(Call c, Set of Program Points pts)
    Add pts to work_queue
    While work_queue not empty
        Let pt = remove point from work queue
        Let st = state associated with pt in c
        Do abstract interpretation of instruction at pt starting at state st
            Result R = set of <next program point, next state> pairs
            If interpretation involves a field (array) set then
                For all references <fcall,fptt> to the field (array)
                    If fcall == c then add fpt to the work queue
                    else queue(fcall,fpt)
            If interpretations involves a return then
                If the return value differs from that saved for the call
                    For each call site <ccall,cpoint> of c
                        queue(ccall,cpoint)
            If interpretation involves an invocation then
                Foreach call nc referenced the the invocation
                    add this invocation to nc
                    If this results in a change, queue(nc,start of nc)
                    If the call has a result value, merge it state
        Foreach <npt,nst> in R
            Let ost = state associated with npt in c
            Let mergst = merge or ost,nst
            If (mergst != ost) then
                Save mergst as state associated with npt in c
                Add nst to the workqueue
```

Fig. 2. Main Analysis Pseudo Code

(for flow analysis purposes) all the found tests. This takes into account the appropriate *@Before* and *@After* annotations to ensure the tests are set up correctly.

The heart of the flow analysis is in analyzing a call one instruction at a time as shown in Figure 2. This is done by maintaining a work queue of program points within the called method that need to be analyzed. The code takes the next program point, finds the associated state, and then does an abstract interpretation of the corresponding byte code or abstract syntax tree node.

The general result of this abstract interpretation is a set of next program points and the next state value at those points. Some instructions may have multiple next points, for example, branches where either condition can be true in the given state. Others might have no next point, for example, a field access where the state indicates the base object must be null. For each point and state in this result, the system checks if merging the new state with any existing state at that point yields a change, and, if so, adds the new state to the work queue.

These results reflect the effect of conditionals and other statements that have data implications. For example, if a value is tested and is known on one branch to be non-null, then the program state for that branch will reflect that. Similarly, the values will reflect integer bounds based on conditionals. Also if a value is accessed so that it is known to be non-null, that information will be propagated.

The interpreter has to handle a variety of special cases. For field or array sets, it checks if adding the new value to the field or array changes the value associated with the field or array element, and if so, queues all references to that field. If the references are within this call, the corresponding program point is added to the work queue; otherwise the referencing call and program point are queue in the system work queue.

If the instruction involves a return from the current method, then the interpreter merges the returned value with the return value stored as part of the call. If the result is a new value, it queues any call sites to the current method in the system work queue.

If the instruction involves an invocation of a method, then the interpreter finds all call objects associated with this interpretation. It first merges the arguments with the call by adding this invocation. If this resulted in a changed initial state, it queues the called method with its starting program point on the system queue. If the call has an associated return value, then this value, if not void, is merged into the current state. The set of next states in this case will be empty if no call has returned yet, will be the next instruction if a call has returned, and will also be extended with states representing any exceptions thrown by the calls.

To speed up the flow analysis without compromising accuracy, FAIT uses special sources we call prototypes for particular classes. These objects simulate the effects of the various calls on the source without doing any actual flow analysis. This is currently used for the Java collection classes (subtypes of *Collection* and *Map*) and for string builders (*StringBuilder* and *StringBuffer*). These particular classes were chosen since abstract interpretation of them can be expensive and we could achieve as good or better accuracy though the simulation. The prototype code tracks the internal values associated with the prototype object, for example the contents of the collection or the key and value collections for a map, and returns appropriate values for the various method calls based on this. It also handles changes in the interpretation process, queuing accesses to the structure whenever the returned value can change. The mechanism for implementing prototypes is general, and it is relatively easy to add new classes if necessary.

**Backward Flow.** While abstract interpretation represents a forward flow analysis, there are points where information should also flow backwards to provide more accurate checking. For example, accessing a field based on a value implies that value is non-null. This information can be propagated forwards (as above), but can also be propagated backwards so that the source for that value is implicitly non-null. FAIT includes facilities for doing such back propagation. The analysis can flag a particular reference value (stack or variable slot) with the information that should be propagated from a particular program state. Then for each program state that led to that state, the corresponding location of that value is determined and back propagation continues. Such backward analysis is done by updating the program states associated with a call.

**Detecting Errors.** Problem detection depends on the kind of specification. Type-based specifications are checked when a value is cast, either explicitly or implicitly (as in a call). The restriction rules for a subtype can indicate that if a value V is restricted to a value V', then an error should be generated. These errors are saved in the call object at the program point where they were detected.

Automata-based specifications are checked on transitions. If a transition is flagged as generating an error, then when that transition occurs, an error is created and associated with the program state that cause the transition.

The set of detected errors is returned to the caller once the analysis is complete.

**Soundness.** The analysis is basically sound in that it will track all possible normal executions. It is conservative in that it can generate executions that are not actually possible because of the approximations used in representing values, sources, and in reusing methods at different call points.

There are some possible points where the analysis might be unsound. First, the analysis does not take into account the possibility of run time exceptions at each point where they might occur implicitly. For example, it does not consider that each object reference might throw a null pointer exception or that an enhanced for loop might throw a concurrent modification exception. It does handle exceptions that are explicitly thrown. Second, the analysis within a method special cases non-volatile field values of the *this* object, tracking their value with the method. This analysis will not detect if the value can be changed in another thread. Third, unless the user specifies enough information in the resource files, the system might not handle reflection and dynamic loading of classes correctly.

Finally, soundness depends on the resource files. If the user uses the resource files to indicate that a routine should be ignored, then the analysis is only sound if that routine has only well-prescribed effects on the data flow. For example, it assumes that the only methods called on input objects are benign (e.g. *equals*, *hashCode*, *toString*), and that no internal state is changed. Any other callbacks or side effects would have to be handled explicitly in the resource file.

The accuracy of the security errors flagged by the analysis is dependent on the soundness of the analysis and on the accuracy of the user's supplied annotations and the models used for handling changes to the subtypes and program states.

V. RESOURCE FILES

Tuning the analysis for performance while achieving the necessary accuracy can depend both on the application and the checks to be made. Significant portions of the application or of the included libraries are often irrelevant to the flow analysis for the security problems of interest. For example, code to parse resource files can be assumed to return any valid return structure and need not be considered in detail. Similarly, unless the checking is concerned with graphical output, most Java Swing routines can be effectively ignored.

To handle these application-specific items, FAIT supports a hierarchy of resource files. The initial file is defined by FAIT itself and covers Java library routines that are generally unneeded or can not be interpreted (e.g. native methods). Each user library can have its own resource file which can provide information about that library or override the system file. Then the application can provide its own resource files which override the library and system files. These resource files can be the same as the ones used to define type and automata based conditions.

The resource files are used to specify how routines should be handled. The most frequent use is to specify that the flow of a particular routine should be ignored and a call to that routine should just be assumed to return an arbitrary (mutable) value of its return type with subtypes based on the input parameters. This can be done for individual methods, for all methods of a class, or for all methods of all classes in a package. In the latter cases, it is possible to provide alternatives for specific methods or classes. The file can also specify a specific return type or replace a particular method call with a call to a different method. They can also designate whether a method should be non-context sensitive, context-sensitive on the *this* parameter, or context sensitive on all its parameters. The latter could be used to identify libraries that are important to security analysis and have those libraries analyzed in more depth.

The flow analysis handles the complexities of real Java systems including static initializers, native methods, Java reflection, callbacks, and threads using this resource file as necessary. It will automatically call the static initializer for a class the first time the class is referenced or when indicated in the resource file. Native methods are handled by assuming they return an arbitrary value of the return type unless otherwise specified by a resource file, with the system method *arraycopy* handled internally. Class-based reflection is handled by returning a mutable object in general. This can be made specific in a resource file, for example having the method *FileSystem.getFileSystem* return an instance of a *UnixFileSystem*. Reflective constructor calls can invoke a particular set of constructors. Method based reflection is handled by having the resource file note the method(s) that could be called at that point.

Java programs often register callback objects, for example in setting up user interface routines through Swing. In some instances (for example, Swing), these routines are only called based on actions that cannot be seen by the analysis. The resource files let the user specify a particular call as registering a callback and will then automatically invoke that callback with appropriate arguments. Finally, threads are handled by having the resource file map *thread.start*() into a separate invocation of the *run* method of that thread.

## VI. INCREMENTAL UPDATE

Incremental update is handled at the file level. The system maintains a list of files that have changed based on edit notifications it receives from the environment. It also listens for messages indicating errors in the source. When it determines that there are no errors and that files have changed, it triggers an incremental update.

The incremental update first updates any resource files that have changed and then ensures that all files are compiled consistently by building new abstract syntax trees as needed and then recomputing all binding information. It then maps any information in the existing flow analysis to use the new abstract syntax trees and binding information.

Updating the flow analysis itself is shown in Figure 3. It begins by removing all flow information associated with the

```
Incremental Update
    Let UPD = set of all calls based on files to be updated
    Let FLDUPD = set of all fields defined in files to be updated
    Let SRCUPD = {}
    Let CALLUPD = { }
    Let SETUPD = { }
    Let VALUPD = { }
    Foreach call C in UPD
        Remove C from set of calls
        Foreach source S defined in C, add S to SRCUPD
        Foreach call C' invoked from C
            Remove C as caller to C'
            If C' has no callers, add to UPD
        Foreach caller site C' of C
            Add C' to CALLUPD
    Foreach source S in SRCUPD
        Remove the source
    Foreach entity sets ES
        If the entity set contains any source in SRCUPD
            Find the entity set NES = ES - SRCUPD
            Add ES -> NES to SETUPD
    Foreach value V
        Let ES = entity set of V
        Let NES = SETUPD(ES)
        If NES is defined, then
            Let NVAL = value based on NES
            Add VAl->NVAL to VALUPD
    Foreach state, field, array and prototype
        Replace internal values with VALUPD values if needed
    Foreach class C that have been statically initialized
        If C is in update set then
            Indicate C hasn't been initialized
            Queue static initializer of C
    Remove all calls in UPD from system work queue
    Queue all calls in CALLUPD
    Foreach fields F in FLDUPD
        Queue all accesses to F
```

Fig. 3. Incremental Update Algorithm

updated files. Then it queues all locations that might be affected by the changed files. Finally it uses the abstract interpretation framework to update the flow until nothing changes.

The update works once for all changed files. It first computes the set of calls that should be removed. This includes all calls to methods in classes in the set of changed files as well as all calls that are only invoked by calls that are being removed. Next it removes all sources that were created in any of these removed calls. This involves removing the source from every source set it is contained in and results in a mapping of each source set to its updated set. It uses this mapping to update all saved values. These can be in fields, arrays, prototypes and states.

Finally the updater updates the flow analysis controller. If any of the classes defined in the changed files were initialized and have a static initializer, it notes that the class is no longer initialized and adds the static initializer to the system work queue. It adds all call sites of any removed call to the system work queue. Finally, it adds any references to fields in the updated files to the work queue.

This process ensures that all methods in the updated files that were invoked previously are checked again and the new information is propagated. It also ensures that all internal tables are links are updated. If new entities are created, they will be propagated and handled by the reanalysis.

The incremental update time is linear in the size of the flow analysis data structures since internal data structures containing calls, source sets, and values are each scanned once. If this becomes too slow on larger systems, it could be reduced at the cost of some tracking overhead.

We note that, while the incremental analysis is still sound, it can yield slightly different results than a complete reanalysis. The incremental update may be more conservative than the full analysis since stored values associated with fields and prototypes are not fully recomputed and contain information associated with the old analysis. It might be more accurate in that method calls that were previously merged, might yield a call that is analyzed separately.

## VII. Providing Problem Explanations

In order to provide information about the potential security problems that flow analysis identifies, the system needs to provide users with details as to why a problem was flagged. These details can be viewed as a graph that shows a sequence of events such as calls, variable assignments, loads, stores, and operations, that lead from a relevant starting point to the point of the error. The easiest way of computing such a graph from the flow analysis is to do a backward analysis from the point of the error to a valid starting point.

This backward analysis is done one step at a time. At each step there is a context describing what is relevant, a program point, and a graph node. The program point is used to determine the flow analysis state. Then all states that are immediate predecessors of this state are considered and an appropriate new context is created for the prior state if possible. If no new context can be created, then the current graph node is checked to see if it represents a valid starting point for the analysis and is marked as such if so. As part of creating the new context, a new graph node may be created linked to the current node. This is done in cases where something relevant has changed, e.g. a value was loaded onto the stack from a variable, a call or return was performed, or a computation was done.

What is relevant to the problem in a context and the computation of the prior context depends on the type of problem. For a type-based error, the context includes the location of the value (stack location, local variable, or field), the subtype causing the error, and the subtype value of the error. For example, it might note that the top of the stack contains a value for SQL checking that is *SqlTainted*. The previous context determines where that value was based on the operation performed between the previous state and the current one and determines if the value from the previous state is relevant to the problem. This might change the associated subtype value, for example changing *SqlFullyTainted* to *SqlTainted*.

For a state-based error, the context consists of the state condition being checked and the state value for that condition. The previous state is checked to see if it has the same state value associated with the state condition. If it had a different value and the value was not the default value for the condition, a new next state is created based on the different value. If the different value is the default value for the condition, the current state is considered a valid starting point for the graph.

The actual algorithm is shown in Section Fig. 4.. The analysis itself is a work queue algorithm. The computation of the prior state is handled in *ComputeNext* and *HandleNext*. *Com-*

```
SafetyContext:
    Fields:
        Safety condition
        Safety condition state

SubtypeContext:
    Fields:
        Subtype
        Subtype value
        Reference to value

BackwardAnalyze(error)
    WORKQUEUE = new Queue<ProgramPoint,Context,Graph>();
    Let ctx = Build context for error
    Let ppt = Program point of error
    Let graph = initial graph node for error
    Add <ppt,ctx,graph> to WORKQUEUE
    While WORKQUEUE not empty
        Let PPT,CTX = remove top of WORKQUEUE
        ComputeNext(ppt,ctx,graph)

ComputeNext(ProgramPoint ppt,Context ctx,Graph graph)
    Let curstate = program state associated with ppt
    If ppt is at the start of a method then
        Let priorctx = context for a call based on ctx
        Let g1 = Add graph node for method entry to graph
        Foreach call site of this call
            Let st0 = program state at call site
            If st0 is relevant to priorctx then
                Let ngraph = Add graph node for call to g1
                Queue <ppt,priorctx,ngraph>
    Else
        Foreach prior state st0 of curstate,graph
            HandleNext(st0,curstate,ppt,ctx)

HandleNext(State st0,State curst,ProgramPoint ppt,Context ctx,Graph g)
    Let priorppt = program point for st0
    Compute the effect of the code at st0 going to curst
        Yields priorctx : context at st0
        Yields auxppts: List of program points of references
    Foreach ppt in auxppts
        Let st1 = state at ppt
        Let auxctx = context for reference
        If st1 is relevant to ctx then
            Let g1 = Add graph node for store to g
            Queue <ppt,auxctx,g1>
    If priorctx is relevant to context then
        Queue <priorppt,priorctx,g>
    Else If priorppt is a method call
        Let retctx = context for a return based on ctx
        Foreach call pcall invoked form priorppt
            If the call return is relevant to ctx
                Foreach return state st1 in pcall
                    If st1 is relevant to ctx then
                        Let retloc = program point for st1
                        Let g 2= Add graph node for return to g
                        Queue <retloc,retctx,g2>
    Else If priorctx is an automata state transition then
        Let g3 = Add graph node for state transition to g
        If priorctx is not a ground state for automata then
            Let pctx = context for prior state
            Queue <priorppt,pctx,g3>
    Else
        Mark graph node g as starting point
```

Fig. 4. Backward Analysis Algorithm

*puteNext* deals with the start of a method, continuing the analysis at each of these that is relevant. *HandleNext* handles transitions inside a method.

The bulk of the implementation is in dealing with the effect of the code going from a program point to a predecessor. This computes the prior context as well as a set of auxiliary program points. The prior context includes an updated reference for subtypes; The auxiliary points are used for variable, field, array, or prototype sets if the instruction accessed one of these. Operations such as string concatenation, where there are multiple potential prior references, will add the additional operands as auxiliary points. Unanalyzed method calls are treated as operations based on their parameters.

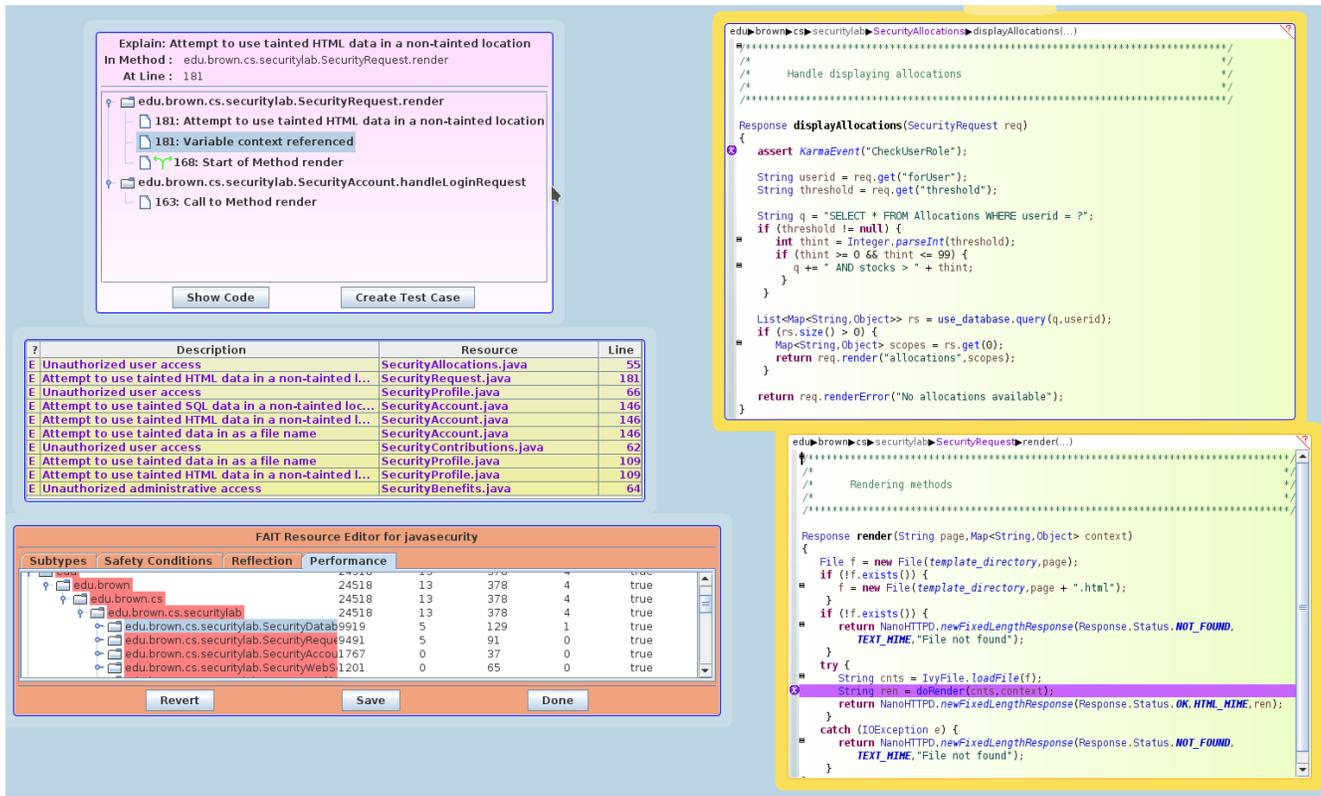

Fig. 5. Annotations showing problems and the problem table view.

When the previous state is the return from a call, the back flow determines if the call was relevant or not (i.e. if the error context can be found without considering the call). If the call is not relevant, it is ignored. If it is relevant, a prior state is set up for all returns from all routines that were called at that point that would be relevant. When the current state is the start of a method the back flow proceeds to each call site of that method where the arguments or program state are relevant.

The context is used to determine if a prior state is relevant. For a safety condition, this means it has the same automata state. For a subtype condition, this means that the referenced variable has the indicated value for the specified subtype.

The system integrates this facility with the environment by allowing problem-specific queries to be sent as messages. It sends back a full description of the resultant graph.

## VIII. INTEGRATING INTO CODE BUBBLES

To demonstrate that FAIT can be used within a development environment and used as the user types, we integrated the initial implementation into the Code Bubbles development environment. [4]. Code Bubbles is a working-set centric integrated development environment designed to let the programmer easily view a complete working set (all the code needed for a single task) and to simplify navigation. It is designed to make it easy to integrate new tools [35], and provides all the hooks needed to integrate a facility similar to what we needed.

FAIT runs as a separate process, communicating with Code Bubbles through messages. FAIT listens for messages regarding new editors, file changes, and error messages; it queries the environment to get information about the current project, the class path, and the current contents of edited files. FAIT listens to Code Bubbles for commands to start the analysis, to indicate what files the user has open for editing, and to provide information about a particular error. It sends the results of the analysis asynchronously to the front end when it becomes available.

If FAIT is enabled in Code Bubbles, it is automatically started if the project uses the Karma library or standard Java type annotations (annotations.jar). FAIT runs the initial analysis and then waits for file changes or new files being opened. After any such change, it waits until there are no compilation errors and then starts an incremental analysis. If a change occurs while analysis is being done, the analysis is suspended so that it can be resumed as part of the next incremental update.

To check the results of the FAIT analysis, we added a preliminary user interface to Code Bubbles for showing and exploring security problems. The user can bring up a window that contains a table of detected problems which can be sorted by problem name, file, or severity. An example of his window is shown in the bottom left of Figure 5. The table can be used to go to the particular source file.

In addition, an annotation is created for problem lines in any editor displaying the corresponding source. Such annotations can be seen as the purple circles in the annotation bar of the editors on the right side of Figure 5. Hovering over these annotations will bring up a short description of the error.

From either the annotation or the error table, the user can request an explanation of the problem. This uses the query mechanism of FAIT to return the problem graph. The graph is displayed one path at a time in a simple interface that shows where there are alternative paths and lets the user explore such paths. An example of this is shown in the upper left of the figure. The user can click on the lines here to bring up the corresponding code editor. In addition, when selecting or moving through the lines here, the corresponding editor locations will be highlighted. Clicking on the fork icon in the display will show the next alternative path in the flow graph at that point.

The explanation shown in the figure is for the use of tainted HTML data in the routine *render* (shown at the lower right). The error is detected at the call to *doRender*. The explanation notes that this is due to the variable *context* being tainted. The next line of the explanation indicates that *context* was passed in as a parameter in a call from *displayAllocations* (upper right window) at line 71. The offending second parameter was then set at line 70 using the result of a database query. This is a potential problem since the data stored in the database is not assumed to be HTML sanitized.

In addition, we created an interface for viewing and editing the resource files. This is shown in the bottom right of the figure. The interface contains four tabs. The first lets the user view, edit and create subtypes. The second lets the use view, edit, and create automata-based safety conditions. The third identifies instances of Java reflection in the flow analysis and lets the user customize these. The final one, shown in the figure, provides performance information to let the user tune the analysis to the application. This includes information (in the last column) as to whether this routine or class seems important to the security analysis.

## IX. EVALUATION

To test our flow analysis and determine if it is practical to integrate into a development environment so it can be run as the user edits, we did a range of experiments using the Securibench test suite from Stanford [25]. As distributed, this consists of seven Java projects that are serlvet-based web applications with some security problems. In addition, we used the insecure web server from our web course described earlier.

**Setup.** The code in Securibench is relatively old, designed for Java 1.4 rather than the current versions of Java. We first updated the code so that it compiled under Java 10 without any compilation errors, mainly by using generic types and replacing deprecated calls, but also by fixing obvious bugs in places and moving to the newer version of JUnit. We then created a library of mocking classes to be used in place of Tomcat or a similar Java web framework. This library includes 11 classes that handle input, output, request data, response data, configuration, and sessions.

Next we added security annotations in the resource file for the mocking library that marked calls to SQL query routines as non-SQL-tainted (12 annotations); that marked output to *ServletOutputStream* or *JspWriter* as non-HTML-tainted (10 anno-tations); and that marked the various calls to open files as non-FILE-tainted (7 annotations). These were added in the resource file for the mocking library. We also marked the calls returning information a malicious user has control over in the mocking library as tainted (24 annotations). No annotations were added directly to the individual projects. We also considered our insecure web server described earlier. This did not use the mocking framework and included 8 annotations and 7 event triggers in the actual code.

TABLE I: PROJECT INFORMATION

| Project | LOS | #File | #Classes | Resource File |
|---|---|---|---|---|
| blueblog | 4,434 | 32 | 342 | 37 |
| jboard | 10,375 | 68 | 1,549 | 32 |
| pebble | 43,092 | 334 | 1,317 | 129 |
| personalblog | 6,219 | 38 | 1,292 | 16 |
| roller | 48,292 | 233 | 2,230 | 46 |
| snipsnap | 48,199 | 438 | 2,076 | 298 |
| webgoat | 8,715 | 36 | 300 | 29 |
| javasecurity | 2,065 | 12 | 1,190 | 73 |

Securibench introduced a dummy class *InvokeServlets* for flow analysis that called some of the available servlets. This is designed to take the place of a web server framework such as Tomcat that invokes servlets using reflection and dynamic class loading. We found the provided dummy classes to be insufficient in that a) they did not include all servlets or JSP calls; and b) they did not make the appropriate initialization calls to set up the servlets. Our analysis is sensitive enough so that if the initialization calls were not invoked, many of the flows leading to potential errors were considered impossible.

To deal with these problems, we created a small package that scans the configuration files for the project and creates our own dummy class. This class invokes all user servlets and related calls (actions and JSP tags) after doing the appropriate initializations. We noted that the while in most cases this increased coverage, there was at least one case (jboard) where the configuration files did not include classes that Securibench analyzed by having explicit calls in *InvokeServlets* that did not have a corresponding configuration file entry.

Next we created a resource file for each of the Securibench projects. This first enabled SQL, HTML, and file taint checking. Next it dealt with uses of Java reflection in the packages, noting which objects were created implicitly by calls to *newInstance* or other routines. Where we found sanitization methods, we added annotations for them. Finally we added a small amount of tuning for external libraries (we explicitly did no tuning for files in the project itself), mainly telling the system to ignore XML parsing (Apache Xerces libraries) and logging (Apache logging library).

A summary of the projects is given in Table I. The first seven projects are from Securibench; the project *javasecurity* is our insecure web server. The second column indicates the number of lines of source in the project, while the third is the number of files. The fourth the number of classes analyzed. Since the number of source classes is approximately the same as the number of files, this shows that libraries constitute the

*10*

TABLE II: INITIAL FLOW ANALYSIS EXPERIMENTAL RESULTS

| Project | #Problems | False Positives | #Methods | #Instructions | Compile Time (seconds) | Analysis Time (in seconds) | | | | Query Time (ms) | Graph Size (avg/max) |
|---|---|---|---|---|---|---|---|---|---|---|---|
| | | | | | | 1 | 2 | 4 | 8 | | |
| blueblog | 4 | 0 | 2,988 | 65,489 | 5.7 | 3.3 | 2.6 | 2.1 | 2.2 | 43 | 21 / 41 |
| jboard | 0 | 0 | 89,671 | 1,308,551 | 8.1 | 58.3 | 36.4 | 24.0 | 19.1 | n/a | n/a |
| pebble | 8 | 0 | 27,619 | 621,974 | 10.6 | 19.9 | 13.4 | 10.6 | 9.4 | 18 | 27 / 51 |
| personalblog | 5 | 0 | 66,697 | 1,196,427 | 7.8 | 56.8 | 35.9 | 25.0 | 21.0 | 19 | 13 / 39 |
| roller | 11 | 1 | 154,589 | 2,946,895 | 11.4 | 112.4 | 71.3 | 51.3 | 46.3 | 174 | 945 / 9870 |
| snipsnap | 3 | 1 | 161,586 | 5,481,878 | 10.8 | 154.2 | 90.1 | 67.0 | 52.2 | 540 | 6225 / 18652 |
| webgoat | 9 | 0 | 31,415 | 1,488,232 | 5.5 | 30.0 | 18.5 | 11.9 | 10.4 | 13 | 10 / 14 |
| javasecurity | 10 | 0 | 3,766 | 91,971 | 6.4 | 4.7 | 3.4 | 3.1 | 2.7 | 26 | 28/ 107 |

bulk of the code to be analyzed and hence that the number of lines of source is an under approximation to the complexity of the projects. The last column gives the size of the created XML resource file in lines. The snipsnap project uses a lot of Java reflection, hence its larger size.

We then ran the flow analysis on the various projects using a JUnit test framework that ran Code Bubbles in background, started up FAIT on the project, and then sent the appropriate messages to FAIT to open all source files and invoke the analysis. The experiments were run on a workstation with 2X AMD Opteron 6140 8 core 2.6ghz processors and 128 G of memory. (None of the runs required more than 8G of memory.)

**Flow Analysis.** We did an initial flow analysis on each of the eight projects, determining the run time with 1, 2, 4, and 8 threads, counting the number of problems that were detected, and analyzing those problems.

The results are shown in Table II. The second column lists the number of problems found. We looked at each reported problem to determine if it was justified by the code and the initial set of annotations and represented a potential problem. Two of these (one from roller and one from snipsnap) were obvious non-problems. The rest represent possible security errors. This yields a precision of 95% (38/40) on the Securibench projects. The original Securibench study flagged an error as a source-sink combination. FAIT only flags the sink as the error, leaving the finding of sources to problem explanation. We checked that the sinks for all errors reported in the original study corresponded to errors reported by FAIT. Thus our tool is at least as accurate as Securibench. Other differences between the two studies include that the security annotations are slightly different, minor fixes have been made to the code, and the initial invocations are different.

For *javasecurity*, which we were familiar with, we verified that all the security issues we were aware of related to the problems that we explicitly addressed (SQL, HTML, files, and authentication) were flagged. We also verified that all the problems flagged were actual problems.

The third and fourth columns give an indication of the work done by the flow analysis. The third column is the number of times the analysis of any method was started; the fourth is the number of abstract instructions executed by the flow analysis. These could be actual byte code instructions or visits (initial or backtracking) to an abstract syntax tree node.

Analyzing the performance data from the analyses, we noted that there were potential areas of improvement. One involved the *hibernate* library to provide a level of persistent objects. *jboard*, *personalblog*, and *roller* made use of this library, and at least 10% of the analysis was spent on it. If this were deemed not relevant to the security issues at hand or could be summarized, it could be eliminated. For *snipsnap*, much of the analysis came involved the deserialization code. For *webgoat*, over 60% of the time was spent in one routine (*Screen.setup*), and in the HTML setup methods it called.

The fifth column is the initial compile time needed. This is a worst case time where we are compiling all the files in the project. This is unrealistic for the larger projects in that users will rarely have hundreds of files actively open at one time. This time also includes the cost of loading and analyzing binary files, a one time cost that can be significant.

The next four columns show the time to actually complete the full initial analysis in seconds with 1, 2, 4, and 8 threads. These show that using multiple threads is somewhat effective, especially for the larger analyses, although that effectiveness drops off after 4 threads.

Finally, the remaining two columns provide information about querying for explanations of all the problems found. The first shows the average time in milliseconds needed to process the query. The second is the size of the returned graph, providing the average size as well as the maximum size. The average query time is well under a second. The graph size varies widely. The larger graphs typically occur when a problem occurs in a library routine, has multiple sources, and the flow involves fields or collections. By looking at these large graphs, one can find the critical user routines and annotate them. This will create smaller, more understandable graphs. We did this for *personalblog*. The original graph had over 8,000 nodes and showed an error in a library routine, but all paths passed tainted data through a call to *SessionImpl.find*. We annotated this call and found the same problems with 5 much smaller graphs. We note that while this annotation simplified the resultant graph, it did not change the underlying flow analysis.

**Incremental Analysis.** To measure the cost of incremental analysis, we measured the run time to update the analysis after a single source file was changed. Since most edits will only affect a single source file, this test should be representative of incremental changes. Moreover, since we completely recom-pile and reevaluate the file, what changed does not matter. Where the project had included more than 50 files, we randomly selected a subset of about 50 files for the test (choosing a file with probability 50/N). Note that we did not actually change the file, but only told FAIT it had changed, which was equivalent.

TABLE III: INCREMENTAL ANALYSIS EXPERIMENTAL RESULTS

| Project | Initial Time (ms) | Compile Time (ms) | Update Time (ms) | Incremental Timing (4 threads) (in ms) | | | | Total Avg. Update time (ms) |
|---|---|---|---|---|---|---|---|---|
| | | | | Average | Median | Min | Max | |
| blueblog | 2,142 | 175 | 126 | 55 | 37 | 30 | 155 | 356 |
| jboard | 24,003 | 275 | 1,484 | 2,334 | 45 | 31 | 9,822 | 4,093 |
| pebble | 10,583 | 958 | 1,124 | 191 | 47 | 37 | 806 | 2,273 |
| personalblog | 24,974 | 223 | 1,091 | 2,928 | 69 | 31 | 9,226 | 4,242 |
| roller | 51,315 | 1,434 | 4,397 | 6,613 | 172 | 44 | 23,691 | 12,444 |
| snipsnap | 66,976 | 1,408 | 4,172 | 3,455 | 59 | 49 | 12,169 | 9,035 |
| webgoat | 11,896 | 236 | 767 | 318 | 35 | 28 | 3,560 | 1,321 |
| javasecurity | 3,123 | 97 | 156 | 74 | 61 | 33 | 189 | 327 |

When running the test, we also made sure that the reported errors remained the same. The results of these experiments are shown in Table III.

The second column is the initial analysis time in milliseconds (taken from the prior table). The next two columns show the average time (in milliseconds) of recompiling after a change and then updating the existing data flow to reflect the changes needed by incorporating the recompilation and removing all items from the changed files. The compilation times are significantly less than those for the initial compilation, but still reflect having to compile all the sources rather than the ones the user might have open.

The update times are dominated by the size of the original data flow analysis data structures since several of these structures are scanned as part of the update as noted in Section VI. These times will vary slightly with the size of the update or how wide-ranging the effects of the edited files are in analysis, but not significantly. We note that there is considerable room for improvement in the update algorithm.

The next columns show the timings for the incremental analysis, providing the average, median, min and max times in milliseconds. These times vary widely. The low values for the median time show that, for the majority of the files, the analysis is almost immediate. (The total time is be dominated by the compilation and flow update time.) Some files, however, can take considerable time (about one-half of the full analysis time). Typically these are files that provide a starting point for potential flows that then have to be reconsidered throughout the analysis.

The total update time (compilation + update + analysis) for the various applications, shown in the last column, varies from well under a second to almost 13 seconds for the more complex applications. These update times would be sufficient for incremental update, although the times for larger applications are currently marginal for immediate feedback.

There are some obvious ways that these times could be improved. First, better tuning of the analysis through the application-specific resource file would yield a smaller analysis and hence faster update and analysis times. Second, the incremental update algorithm could be modified to avoid scanning the whole data structure and only scan what was changed. While this would complicate the update algorithm, much of the dependency information is already available to do this.

Third, update could be done at the method level rather than the file level when possible. This would require additional bookkeeping to determine what methods changed, but would otherwise fit into the incremental analysis framework and would reduce the actual analysis time.

Fourth, the bulk of the reevaluation time, especially for the longer runs, involves tracking sources created in the edited code. The original sources are removed wherever they occurred, and the flow for any new sources to all of these locations needs to be recomputed. It might be possible to reuse the original sources which would limit the update to places where the types or subtypes associated with those source change.

**Comparison of Results.** There are other tools available to do the types of analysis that FAIT accomplishes. The two primary ones are the Checker framework and the CHEETAH system.

As we have noted, Checker is able to do the subtype checking that FAIT does. However, doing it for these examples would have required substantially more type annotations since one would need to annotate all the locations that a value might flow to within the code. We were able to do this by simply annotating the mocking library.

CHEETAH does a just-in-time flow analysis, so it can produce results quite similar to ours. However, it currently does only taint analysis and doesn't handle general subtypes or automata conditions. Our incremental update times to produce full results are about the same as CHEETAH uses to provide its initial, more general, results. Our initial computation times are generally less than CHEETAH's time for its full analysis. Note that this is not a direct comparison as CHEETAH was run on different systems,

**Threats to Validity.** There are several threats to the validity and applicability of these experiments. First, the projects might not be representative of modern applications. The Securibench projects are all web applications and are 10+ years old. We tried to analyze the current version of Webgoat, but it uses

Java bean processing to effectively add or modify the code at run time using reflection, a capability that makes it difficult for any static analysis to work.

Other types of applications might have different behaviors. We have run the analysis (generally without checking for problems) on a variety of systems to validate that it works, it scales, and the results seem to be correct. For example, working on FAIT itself involves a set of projects totaling 130KLOS. The initial evaluation takes 20-30 seconds, but the incremental updates are typically done in 1-2 seconds. Certain items, such as parsers seem to require much more analysis time than other types of code, but are easy to deal with in resource files.

Next, the flow analysis involves both randomness in the choice of what method to work on next and multiple threads. This means that runs on the same source will be slightly different in terms of effort and running time even though they generate the same result. The variation seems to be less than 1%, so the numbers are probably accurate, but there is always the possibility of a particularly bad ordering that causes excessive recomputation. Also, since the flow analyzers for byte and source code are different, different combinations of source files might yield different timings. Our experiments were done using source for all the project files. Finally, randomness was involved in selecting which files were updated.

We did minimal tuning in the resource files for the particular applications. In actual usage, we would expect that more tuning would be done and hence the timings could be significantly smaller. In particular, it should be possible to avoid doing a full analysis of libraries such as hibernate or irrelevant aspects of the actual code. Our experience is that we can cut the initial analysis time in half with minimal specifications and will also reduce update times.

We have done minimal performance tuning of FAIT, especially for multiple threads, and expect that we can reduce the timings by doing so. Finding and eliminating synchronization bottlenecks could increase the effectiveness of using multiple threads. Finally, we note that the code still contains bugs which could affect both the performance and accuracy.

**Limitations.** While FAIT is a good first step toward continuous flow analysis, it has its limitations. The first is that it requires the developer to create a resource file in order to be effective. This might be needed to achieve the desired performance, and is needed to handle any uses of reflection in the system, as well as to define the security properties of interest. Additional resource files might have to be created for the various libraries the system uses, although these can be part of the developers resource file.

Second, FAIT does not handle the latest Java capabilities. Because we are using our own compiler to create annotated abstract syntax trees for code that is actively edited, FAIT can only handle constructs we can compile. We currently handle all of Java 10. FAIT also cannot easily handle programs that make extensive use of reflection to effectively change the code at run time.

Third, FAIT can require significant CPU and memory. Although it runs in background and we are generally using it with 4 threads, it can make use of all those threads and has been using 4-8 G of memory on larger runs. A developer would need to have a powerful workstation or set it up to run on a separate machine.

Finally, as we noted, the explanations can be long and have many paths. It can be difficult to understand or explore an explanation with thousands of nodes. The developer can get around this by adding appropriate annotations to the code or resource files as we did for *pesonalblog*, but that is more work.

## X. CONCLUSION

We have developed a flow analyzer aimed at identifying security problems that achieves a balance between performance and accuracy. It is designed to run incrementally as the user edits, providing almost immediate feedback on potential security problems. It uses flow analysis to minimize the number of annotations needed to define potential security issues. It uses a resource file to tune the analysis and to define the particular security problems of interest. It can provide detailed explanations of any problems it finds. An initial version of the tool is integrated into the Code Bubbles development environment.

The system is available. The source for the analyzer, the mocking framework, and the karma library are available on GitHub. The current version of Code Bubbles on SourceForge includes the code for integrating with FAIT. Our version of the Securibench projects and our javasecurity project as well as the scripts used to run the various experiments are available from our ftp site.

Finally, we note that while FAIT is currently running with Code Bubbles, it is designed to be independent of the actual programming environment. It runs as a separate process and uses asynchronous messages to communicate with the environment. Another environment could easily be made to send similar messages where needed and thus use the framework without modification.